# Coherent Coupling of WS$_2$ Monolayers with Metallic Photonic Nanostructures at Room Temperature

Shaojun Wang,[†,#] Songlin Li,[†,#] Thibault Chervy,[†,#] Atef Shalabney,[†,‡] Stefano Azzini,[†] Emanuele Orgiu,[†] James A. Hutchison,[†] Cyriaque Genet,[†] Paolo Samorì,[†] and Thomas W. Ebbesen[†*]

[†] ISIS & icFRC, University of Strasbourg and CNRS, Strasbourg 67000, France

[‡] Braude College, Snunit St 51, Karmiel 2161002, Israel

[#]These authors contribute equally for this work.

**ABSTRACT**：Room temperature strong coupling of WS$_2$ monolayer exciton transitions to metallic Fabry-Perot and plasmonic optical cavities is demonstrated. A Rabi splitting of 101 meV is observed for the Fabry-Perot cavity, more than double those reported to date in other 2D materials. The enhanced magnitude and visibility of WS$_2$ monolayer strong coupling is attributed to the larger absorption coefficient, the narrower linewidth of the *A* exciton transition, and greater spin-orbit coupling. For WS$_2$ coupled to plasmonic arrays, the Rabi splitting still reaches 60 meV despite the less favorable coupling conditions, and displays interesting photoluminescence features. The unambiguous signature of WS$_2$ monolayer strong coupling in easily fabricated metallic resonators at room temperature suggests many possibilities for combining light-matter hybridization with spin and valleytronics.

**KEYWORDS**：*strong coupling, 2D materials, polaritonics, surface plasmons, WS$_2$ monolayers*

The strong coupling of light and semiconductor excitons to form exciton-polaritonic states plays a role in many fascinating recent advances, ranging from low-threshold lasing,[1, 2] and Bose-Einstein Condensation,[3] to enhanced charge transport,[4] workfunction tuning,[5] and phase transition modification.[6] These advances draw on the quasi-bosonic nature of exciton-polaritons and the dispersive and delocalized nature of the hybrid states. The strong coupling limit is reached when coherent energy exchange between the excitonic transition and a resonant optical cavity overcomes other relaxation pathways, the spectral signature of which is the splitting of the absorption band corresponding to the two polaritonic states.[7-9] The splitting of these states at resonance, *i.e.* the Rabi splitting, $\hbar\Omega_R$, measures the coupling strength and depends on the scalar product of the electric field per photon ***E*** in the cavity and the exciton transition dipole moment ***d***.

The choice of appropriate optical cavity and semiconductor transition always involves a compromise between maximizing the ***E·d*** product while minimizing losses. On the photonic side, this typically lies in the choice of either distributed Bragg reflectors (DBRs) with high Q factors but diffuse mode volumes, or metallic resonators which concentrate the optical field in sub-wavelength volumes but suffer absorption losses. On the excitonic side, one ideally seeks a highly allowed (direct band gap) dipolar transition in the visible/near-IR spectral region forming a tightly bound energy exciton for room temperature stability, with minimal non-radiative relaxation pathways and inhomogeneous broadening. Monolayer transition metal dichalcogenides (TMDs) exhibit excitonic transitions that satisfy all these criteria.

During the past five years, it has been demonstrated that group VI TMDs with structure $MX_2$ (where *M* is Mo or W and *X* is S or Se) display a transition from indirect to direct band-gap semi-conductors when passing from multilayers to a monolayer.[10-12] The resulting strongly allowed excitonic resonances dominate the visible and near-IR absorption spectra of these systems, having exceptionally large binding energies (>0.3 eV) due to the reduced dielectric screening.[13-15]. Figure 1a shows a top view of the $WS_2$ monolayer. The direct band-gap transition in the monolayer case occurs at the energy-degenerate K (K') points at the edges of the 2D hexagonal Brillouin zone.[16-19] Monolayer TMDs exhibit emission yields approaching unity in the absence of surface trap states,[20] which is important for lasing[21] and other optoelectronic applications.[22-24]

Recently the first reports of strong light-matter coupling with monolayer TMDs appeared in literature, all involving $MoX_2$ monolayers. Menon and co-workers[25] and Tartakovskii and co-workers[26] incorporated $MoS_2$ and $MoSe_2$ respectively into DBR cavities, while Agarwal and co-workers[27] recently studied the coupling of $MoS_2$ to both local and surface lattice modes of metal nanoparticle arrays. These studies show the real potential of TMDs for strong coupling but the observed Rabi splitting were limited by the absorption features of the materials with a maximum reported splitting of 46 meV. In particular in $MoX_2$ monolayers, spin-orbit coupling (SOC) induces the splitting of the excitonic transition by *ca*. 150 meV such that both the so-called *A* and *B* exciton transitions (see Figure 1b) can simultaneously interact with cavity modes complicating the studies of such systems.[27] The $WS_2$ monolayer has the advantage that it presents a much sharper isolated absorption band as can be seen in Figure 1b. In addition it displays an intense photoluminescence (PL) peak at 2.016 eV (Figure 1c). Hence $WS_2$ constitutes a natural choice for light-matter strong coupling.

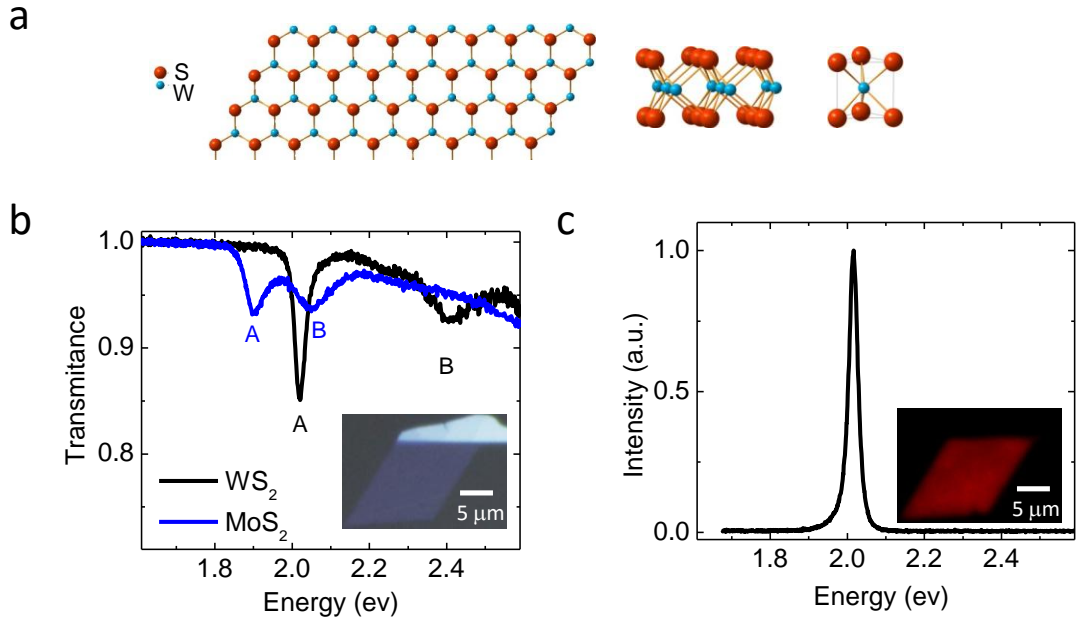

**Figure 1.** (a) The structure of a $WS_2$ monolayer showing, from left to right, out-of-plane view, in-plane view, and the unit cell. (b) Transmittance of monolayers of $WS_2$ (black solid curve) and $MoS_2$ (blue solid curve) on quartz; the inset shows an optical micrograph of the $WS_2$ flake in which the blue region is a monolayer and the bright region is a multilayer. (c) Emission spectrum taken from the monolayer region of $WS_2$ under continuous wave (cw) 532 nm (2.330 eV) excitation. The inset displays the fluorescence image of the $WS_2$ flake in which bright emission is observed only from the monolayer region under cw 470 nm (2.64 eV) excitation.

In this letter we demonstrate that by coupling $WS_2$ monolayers to metallic resonators, the magnitude and visibility of light-monolayer TMD strong coupling at room temperature is substantially enhanced with a Rabi splitting of 101 meV in Fabry-Perot cavities and 60 meV on plasmonic arrays. The energy-momentum dispersion properties of the monolayer $WS_2$ exciton-polaritons are explored by transmission, reflection and photoluminescence (PL) spectroscopy. In particular Rabi splittings in TE and TM dispersion curves give rise to unusual PL behavior. The results are discussed in terms of the potential of coherent light-matter interactions using $WS_2$ monolayers.

To ensure high quality samples and to avoid environmental contamination, the TMD monolayers were exfoliated from bulk single crystals and then dry-transferred onto substrates as described in the Methods section. The *A* exciton of monolayer WS$_2$ absorbs 14% of normally incident unpolarized light, *ca.* ~ 2.5 times more than a monolayer MoS$_2$ (Figure 1). The stronger SOC-induced splitting of the *A* and *B* bands in WS$_2$ (*ca.* 389 meV) is also clear, as is the narrower linewidth of the *A* exciton (28 meV) compared to that of monolayer MoS$_2$ (45 meV) (see table S1 and S2). The WS$_2$ monolayer emission quantum yield, in the absence of special surface treatment methods,[20] was estimated to be 6% by others.[29] The exciton emission (2.016 eV) has a linewidth of 26 meV, displaying a tiny Stokes shift (~ 4 meV) from the *A* exciton absorption. All emission studies herein were conducted at the minimum pump intensity possible (less than 1.0 µW/µm$^2$) such that many body interactions were avoided.[29, 30]

Figure 2a illustrates the generic coupling of a 2D material to resonant optical mode leading to the formation of the polaritonic states P+ and P- separated by the Rabi splitting. $\hbar\Omega_R$ remains finite even in absence of real photons due to interaction with the zero-point energy fluctuations (vacuum field) of the confined electromagnetic field. In this study WS$_2$ monolayers were coupled to two types of metallic resonators: Fabry-Perot cavities and periodic plasmonic structures. Firstly we fabricated Fabry-Perot (FP) cavities with mirrors of 50 nm Ag and the WS$_2$ monolayer was positioned at the field maximum[31] at the centre of the cavity with ± 5 nm accuracy using LiF spacer layers (see schematic, Figure 2b). The cavity thickness was adjusted such that its fundamental mode was tuned to the *A* exciton energy ensuring the smallest mode volume.[31]

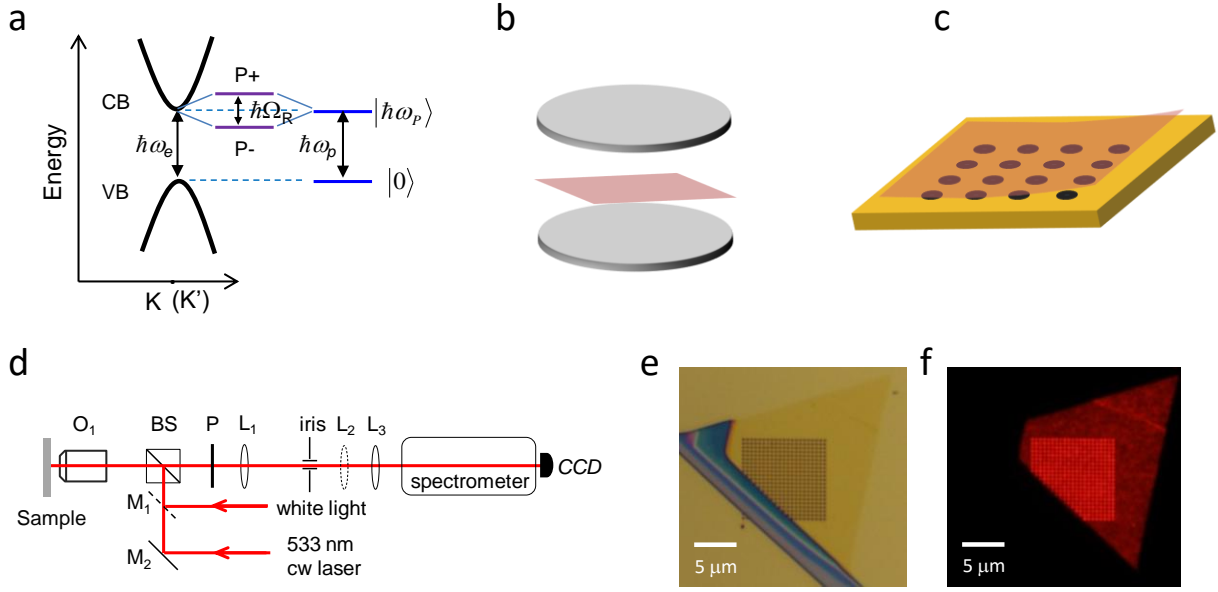

**Figure 2.** (a) Energy diagram of the strong coupling of the direct band-gap transition $\hbar\omega_e$ between the valance band (VB) and conductive band (CB) of monolayer $WS_2$ to the first optical mode $\hbar\omega_p$ of a cavity. The coherent coupling forms two polaritonic states P+ and P− separated by the Rabi energy $\hbar\Omega_R$. (b) Schematic of the Ag Fabry-Perot (FP) cavity with $WS_2$ monolayer placed at the centre (the optical field maximum). (c) Schematic of the plasmonic hole array with $WS_2$ monolayer physisorbed on top. 5 nm of PMMA separates the monolayer from the metal surface. (d) Schematic of the optical setup for angle resolved spectroscopy. White light source or cw 533 nm laser beam is selected by the flipped mirror $M_1$ to pass through a beam splitter (BS) and a 40x objective lens ($O_1$, N.A. = 0.6). The reflected beam or PL from the sample, collected by the same $O_1$, passes through a polarization analyzer (P), tube lens ($L_1$), spatial filtering iris, imaging lens ( formed by $L_2$ and $L_3$), Fourier lens ($L_3$) and spectrometer coupled CCD. (e) Optical micrograph of a $WS_2$ monolayer physisorbed on the plasmonic hole array, the monolayer region is delineated by the red PL image of the same $WS_2$ monolayer in (f).

Secondly a $WS_2$ monolayer was transferred onto hole arrays milled by focused ion beam in a Au film (see schematic in Figure 2c). A 5 nm PMMA film was spin coated onto the array first to avoid emission quenching,[32] or hot electron transfer between Au and the semiconductor,[33, 34] nevertheless the monolayer is still positioned near the plasmonic field maximum at the interface. The surface plasmon (SP) resonance was tuned to the A exciton energy by adjusting the array period[35] (e.g. a period of 530 nm gives a surface plasmon resonance near 2.010 eV when a monolayer flake is on top). An optical micrograph and PL image clearly show the position of the $WS_2$ monolayer on top of a hole array (Figure 2e,f). The emission is enhanced ~ 2.5-fold above the holes possibly due to two factors: firstly, the plasmonic field has a maximum above the holes, enhancing the photonic mode density at this point, thereby increasing the excitonic radiative rate,[32] and secondly the increased dielectric screening where the monolayer is suspended over the hole rather than in Van der Waals contact with the substrate could enhance the emission.[10]

Angle-resolved transmission spectra of the FP cavity with $WS_2$ monolayer are shown in Figure 3a for TE polarization. The progressive dispersion of the cavity mode through the energy of the *A* exciton is accompanied by a clear anti-crossing, which is mapped out in terms of spectral maxima in Figure 3b. After fitting the energy of the two peaks as a function of in-plane momentum $k_{//}$ using the coupled oscillator model (described in the Methods), it is evident that the original exciton transition energy (black horizontal dot-dashed line in Figure 3b) and the cavity mode (black parabolic dot-dashed curve) split into two new bands, P+ and P- (blue dashed curves). Both the experimental data and the fitted curves unambiguously demonstrate an anti-crossing effect, resulting from the coupling between the fundamental cavity mode and the *A* exciton transition. From the fitting a Rabi splitting of 101 meV is extracted. This splitting is greater than the linewidth of the FP cavity mode (~ 80 meV) and the exciton linewidth (28 meV) putting the interaction firmly in the strong coupling regime. The relative photonic and excitonic content of the polariton states can be calculated in terms of mixing coefficients, given in SI, Figure S4 as a function of the in-plane momentum $k_{//}$. The results show that the polaritonic states are 1:1 hybrids of the *A* exciton and the cavity at $|k_{//}| = 4.35$ μm$^{-1}$. The dispersion of the sample also was measured in Fourier space by microscope reflectometry (schematized in Figure 2d) and the results are shown in Figure 3c. The Rabi splitting, again extracted from a coupled oscillator model, was 90 meV, a little smaller than that observed in transmission measurements, as expected.[36] The Fourier image of the photoluminescence from the same cavity is shown in Figure 3d. The emission from the lower branch polariton clearly dominates, but emission is also observed at the uncoupled exciton energy (Stokes shift 4 meV) while upper branch polariton emission is not detected. Notice that the P- emission is centered at $k_{//}$ values where the bare optical mode is iso-energetic with the *A* exciton.

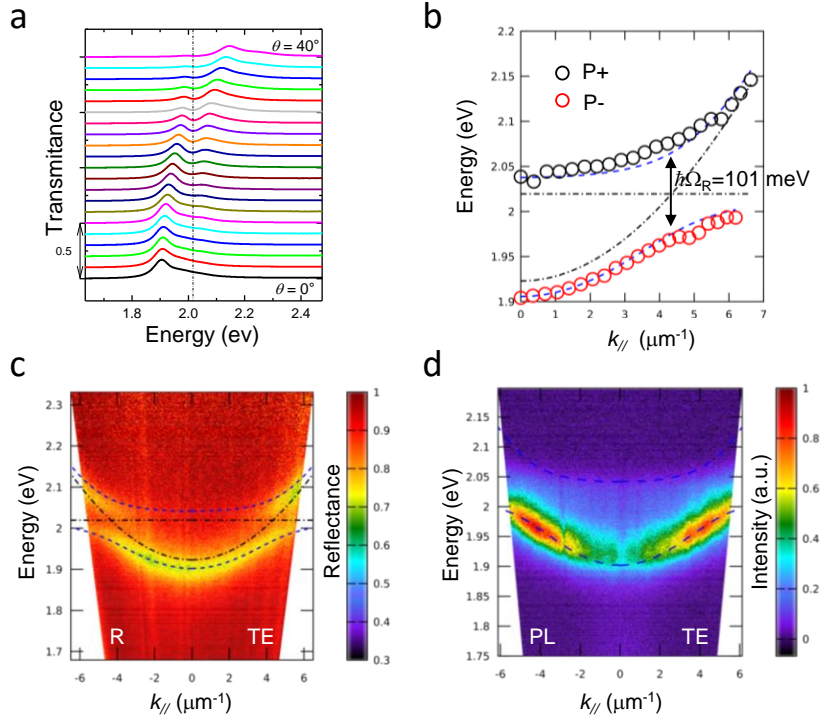

**Figure 3.** (a) Transmission spectra as a function of probe angle (θ from 0° to 40°) for a FP cavity with the monolayer $WS_2$. (b) Peak transmission energies as a function of in-plane angular momentum for the same FP cavity. Black and red circles correspond, respectively, to the measured positions of the P+ and P- extracted from the data displayed in **a**. Black horizontal dot-dashed line and curves show, respectively, the dispersion of the A exciton transition energy and empty cavity mode. The P+ and P- band are fitted by blue dashed curves using coupled oscillator model (see methods) (c) Energy/in-plane momentum dispersion for the same FP cavity taken in TE reflection mode. (d) Dispersion of the photoluminescence from the same FP cavity under cw 532 nm excitation. The energies of P+ and P- bands are indicated, taken from (c).

The dispersion curves of the plasmonic hole array with a monolayer $WS_2$ in both TM and TE modes are shown in Figure 4. Under TM polarization, the anti-crossing between the (0, ±1) SP modes (black dot-dashed curves in Figure 4a) and the *A* exciton (horizontal black dot-dashed line in Figure 4a) is again clear, giving a Rabi splitting of 60 meV. This value is smaller than the one observed for the FP cavity but still easily observable given the width of the SP resonance at 2.010 eV (36 meV). This reduction as compared to the FP cavity splitting can be explained by the relative orientation of *E* and *d* in the two cases. The exciton transition dipole is oriented in-plane in TMD monolayers, thus perfectly aligned with the field polarization of the FP cavity (Figure 2b), whereas for the plasmonic arrays the field is elliptically polarized at the interface, reducing the scalar product of *E* and *d*.

The mixing coefficients for the strong coupling of the (0, -1) TM mode and the exciton in Figure 4b shows that the interaction between them is limited to an in-plane momentum range of -1 to 1 $\mu m^{-1}$ due to the strongly dispersive behavior of the TM mode. This has consequence for the emission as can be seen in Figure 4c. At the anti-crossing, the emission is mainly a mixture of the bare and coupled $WS_2$ (see the solid blue curve in SI, Figure S5 (a)). A weak shoulder can be observed at the position of higher branch polariton compared the emission of uncoupled exciton.

In the TE case, an anti-crossing with a Rabi splitting of 60 meV is also observed for the interaction of the (±1, 0) SP mode with the $WS_2$ monolayer (Figure 4d) even though the quality factor of the bare TE SP mode is a bit lower (Q ~ 20 at 2.010 eV). The dispersion of the TE mode has a parabolic shape as shown in Figure 4d (dot-dashed curves). The slower dispersion of this mode results in a larger range of interacting in-plane momenta, from -5 to 5 $\mu m^{-1}$. In the range -2 to 2 $\mu m^{-1}$, the measured P+ and P- bands are fitted well by the coupled oscillator model. However, beyond this range P+ bends downward due to interactions with the higher (1, 1) SP modes which are not included in the model. The dispersion of P- emission (Figure 4f) also matches well with the predictions of the coupled oscillator model. At the resonant condition, $k_{//} = \pm 1.89$ μm$^{-1}$, the PL spectrum in SI, Figure S5b is dominated by the uncoupled exciton, with weak emission from P+ and P-.

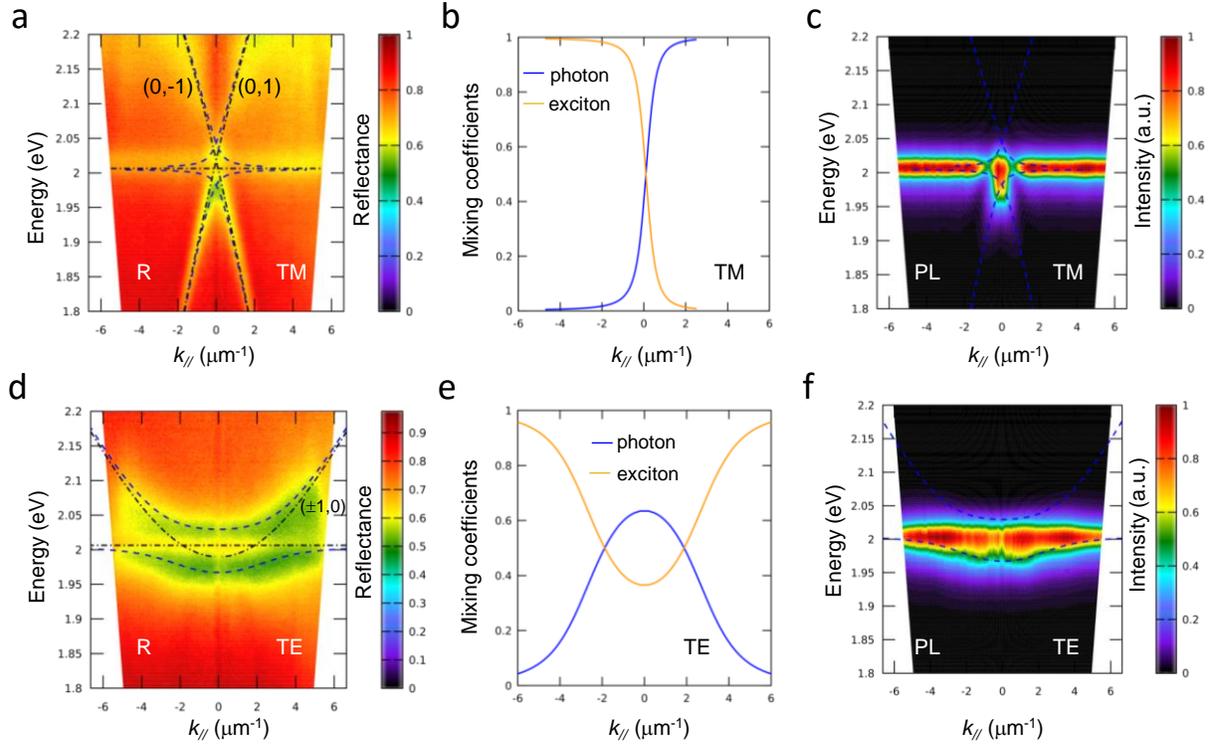

**Figure 4.** (a) and (d) Energy/in-plane momentum dispersion for the plasmonic hole array with WS$_2$ monolayer taken in TM, TE reflection mode respectively. Black horizontal dot-dashed line and curves show, respectively, the dispersion of the A exciton transition energy and empty SP $(0, \pm 1) / (\pm 1, 0)$ mode. The P+ and P- band are fitted by blue dashed curves using coupled oscillator model (see methods). (b) and (e) The mixing coefficients of P- band shown in (a) and (d). Blue and yellow curves represent the photon, exciton content of the P- band respectively. (c) and (f) Dispersion of the photoluminescence from the same plasmonic hole array. The energies of P+ and P- bands are indicated, taken from (a) and (d).

The hint of emission from the upper polariton branch observed here for the plasmonic array (TM polarization) recalls the observation of emission from the upper branch in DBR cavities with MoX$_2$ monolayer.[25,26] Tartakovskii and co-workers[26] point out that the ratio of the Rabi splitting to the exciton binding energy is unusually small in these systems, 0.04 in their case compared to a ratio >0.2 in all other strongly coupled systems ever studied. Therefore, the electron-hole continuum lies far from the polariton resonances and does not efficiently dephase them. Taking literature estimates for the exciton binding energy in WS$_2$ (0.3-0.7 eV),[13-15] the same ratio observed here takes values in between 0.15-0.37 for the FP cavity, and 0.09-0.22 for the plasmonic array. The unusual gaps in the PL under TM polarization near $k_{//}$ =0 most likely originates from the competition between the different modes density in TM and TE that co-exist at this momentum. Another interesting difference between the TM and TE dispersion curves lies in the curvature of the P- branch which concave in the case of TE while sharply convex for TM. This should have dramatic effects on the polariton dynamics in such systems. For instance BEC can only be achieved in the TE case as it requires a stable minimum to accumulate population. Future studies of ultrafast dynamics of these systems will provide more insight into such issues.

In summary we have shown that by incorporating WS$_2$ monolayers into metallic optical nanocavities, clear signatures of strong coupling are observed in transmission, reflection and emission at room temperature. The Rabi splittings observed in WS$_2$ FP cavities are more than double those observed in previous studies on MoX$_2$, irrespective of the type of cavity employed. We attribute these improvements to the superior absorption characteristics of the WS$_2$ *A* exciton in the context of strong coupling which allows unambiguous Rabi splitting to be observed despite the lossy nature of metallic cavities. The latter cavities have many beneficial characteristics, having much smaller mode volumes and being much easier to fabricate than DBR mirror cavities. Perhaps more importantly, plasmonic cavities (either local SPs on nanoparticles[37] or diffractive modes on arrays[38]) are open and thus easily integrated into optoelectronic devices and also accessible for chemical applications. Recent work has demonstrated that a chemical reaction rate can be controlled by excitonic light-matter strong coupling.[39] It has been suggested that such effects can be extended to coupling vibrational modes of specific bonds.[40] Our findings are therefore also particularly relevant given that WS$_2$ and other TMDs are widely exploited materials for applications in catalysis. Finally, the results here suggest that the combination of the already rich spin and valley physics of monolayer TMDs with polariton physics at room temperature should open exciting possibilities in fundamental physics.

**METHODS**

**Sample fabrication.** Atomically thin MoS$_2$ and WS$_2$ films were fabricated by an improved mechanical exfoliation method[41] from a synthetic single crystal (hq graphene, the Netherlands), and the monolayer samples were identified by optical contrast, absorption and PL spectroscopy. The monolayer WS$_2$ flakes were first deposited on a flexible PDMS slab before being transferred onto the substrates (such as quartz, cavity).[6] Beginning with FP metallic cavities, a 50 nm thick silver film was evaporated on a glass substrate, upon which was evaporated an 86 nm thick LiF film. Subsequently, the monolayer WS$_2$ flake was transferred onto this half metallic FP cavity substrate. 90 nm thick LiF and 50 nm thick Ag film were subsequently evaporated onto the flake surface to form a medium quality (Q factor ~ 30 at 614 nm) FP cavity. For plasmonic samples, subwavelength hole arrays were milled by focused ion beam (FIB) in sputtered gold films of 260 nm thickness on a glass substrate covered by a 5 nm thick chromium adhesion layer. The hole diameter and the period of the hole array were 120 nm and 530 nm respectively (shown in the SEM image of Figure S1). To avoid short range interactions between the flake and gold, a 5 nm thick of PMMA film was deposited onto the hole array. Finally, the monolayer flake on a PDMS slab was transferred onto the hole array under microscope.

**Optical measurements.** The transmission spectra of the monolayer flakes on quartz substrate and FP cavity samples were measured using an optical microscope. The samples were aligned along the optical axis of the microscope and illuminated with quasi-collimated white light. The light transmitted by the samples was then collected using a microscope objective lens (20x magnification, N.A. = 0.45) and imaged by a spectrometer (Acton SpectraPro 300i) and silicon charge-couple-device (CCD) (Princeton Instrument VersArray 1300B). The angular dispersion of FP cavity samples was characterized with the transmission goniometric method by rotating the samples from the normal incidence condition to 40° by every 2°. Meanwhile, the reflection and PL spectra of all the samples were measured using a microscope reflectometry setup equipped with an optical Fourier analysis lens ($L_3$ shown in Figure 2d). In this setup, the samples were excited by white light or 533 nm laser beam. A 40x objective (N.A. = 0.60) collected the emission or transmission, directing it to an iris at the focus point of the tube lens ($L_1$). The iris acts as a spatial filter selecting a ~5x5 $\mu m^2$ area of the sample plane. The angular distribution of the reflection or emission from the sample was analyzed in the Fourier plane of $L_3$. A white light beam and lens $L_2$ in the optical path were used as a microscope to locate samples in real space. A linear polarizer was placed in front of $L_1$ to select either TE or TM mode of the cavity samples.

**Dispersion of 2D material calculated by multi-Lorentzian model.** The optical properties of $WS_2$ monolayer was analyzed using a multi-Lorentizian model in order to simulate $WS_2$-cavity interactions using transfer matrices, as described below and in the SI for the details. The findings were systematically compared to the $MoS_2$ system. The refractive index of $WS_2$ extracted from transmission measurements are included in the SI, Figure S2, together with fitted functions from which oscillator strengths and linewidths were extracted (SI, Table S1 and S2). Our estimated dielectric functions are in good qualitative agreement with the literature.[28] The $WS_2$-cavity results (displayed in Figure S3) reveal a Rabi splitting being ~1.51 fold that of $MoS_2$, a value comparable to the ratio of their transition dipole moments.

The dispersion of the $WS_2$ and $MoS_2$ monolayers was retrieved from the transmission measurements using a multi-Lorentzian model.[28, 40] The absorption bands were represented by multiple resonances as $\varepsilon(E) = \varepsilon_B + \sum_{j=1}^{N} \frac{f_j}{E_{0j}^2 - E^2 - iE \cdot \Gamma_j}$, $E$ being the photon energy in eV whereas $\varepsilon_B$, $f_j$, $E_{0j}$, and $\Gamma_j$ are respectively the background dielectric contribution, oscillator strength, resonance energy, and the phenomenological damping constant of absorption band $j$. The absorption intensity of the band is determined by both $f_j$ and $\Gamma_j$, whereas the linewidth is solely governed by $\Gamma_j$. In the fit procedure, all these parameters together were varied to obtain the best fit with the experimental measurements. The thicknesses of the $WS_2$ and $MoS_2$ monolayers were considered as 0.618 nm and 0.646 nm respectively.[28] Since monolayer structures are considered, $\varepsilon_B$ can be reasonably assumed as unity ($\varepsilon_B = 1$). To account for the substrate refractive index, the reflectivity and transmittance of the single interface was calculated from the measured transmission of the bare quartz substrate. The dispersion parameters of monolayer $WS_2$ and $MoS_2$ are given in the table S1 and S2 of SI respectively.

**Dispersion curves fitted by the coupled oscillator models**. The polariton dispersion extracted from the transmittance maxima or the reflectivity minima are fitted by a coupled oscillator model[8] $\begin{bmatrix} E_{ph}(k_{//}) & V \\ V & E_{ex} \end{bmatrix} \begin{pmatrix} \alpha \\ \beta \end{pmatrix} = E_{pol}(k_{//}) \begin{pmatrix} \alpha \\ \beta \end{pmatrix}$, where $E_{ph}(k_{//})$ is the energy of empty photonic mode, $V = \hbar\Omega_R/2$ the interaction potential between photonic mode and exciton with energy $E_{ex}$, and $E_{pol}(k_{//})$ the momentum dependent hybrid eigenvalues of P+ and P- branches. The mixing coefficients $|\alpha|^2$ and $|\beta|^2$ describe the relative photonic and excitonic content of the polaritonic states and $\alpha$, $\beta$ are known as the Hopfield coefficients. For the plasmonic hole array, the dispersion of SP modes are defined by the momentum matching condition as

$$\left|\vec{k}_{spp}\right| = \left|\vec{i}(m\frac{2\pi}{P}) + \vec{j}(n\frac{2\pi}{P} + k_{//})\right| \tag{1}$$

where $\vec{i}$, $\vec{j}$ are unit vectors along horizontal and vertical directions respectively, $k_{//}$ is the in-plane momentum component of the incident light, $P$ is the lattice period, and m and n are integers. The scattering orders of SP modes are denoted by (m, n). From equation (1), the dispersion of degenerated TE ($\pm 1$, 0) mode can be written as

$$\omega = \frac{c}{n_{sp}} \sqrt{k_{//}^2 + (\frac{2\pi}{P})^2} \tag{2}$$

where $\omega$ is the light frequency, $c$ the propagation speed of the light in the vacuum, and $n_{sp}$ the effective index of SP modes ($n_{sp} = \sqrt{\frac{\varepsilon_m \varepsilon_d}{\varepsilon_m + \varepsilon_d}}$, $\varepsilon_m$ and $\varepsilon_d$ are the permittivities of Au film and air).

When $-\frac{2\pi}{P} \leq k_{//} \leq \frac{2\pi}{P}$, the dispersion of TM (0, 1) and (0, -1) mode are respectively described as

$$\omega = \frac{c}{n_{sp}}(k_{//} + \frac{2\pi}{P}) \tag{3}$$

$$\omega = -\frac{c}{n_{sp}}(k_{//} - \frac{2\pi}{P}) \tag{4}$$

## ASSOCIATED CONTENT

**Supporting Information**

Scanning electron microscope (SEM) image of the hole array sample, refractive index and dispersion parameters of monolayer of 2D materials, comparing transmission, reflection of coupled FP cavities simulated by transfer matrix method, mixing coefficients of strong coupled monolayer $WS_2$ with TE mode of FP cavity, and PL spectra of strong coupled monolayer $WS_2$ with TM/TE mode of hole array are attached in the supporting information (SI).

## AUTHOR INFORMATION

**Corresponding Author**

*E-mail: ebbesen@unistra.fr

**Notes**

The authors declare no competing financial interest.

## ACKNOWLEDGEMENTS

This work was supported by the European Commission through the Graphene Flagship (GA-604391) and the FET project UPGRADE (grant no. 309056), the ANR Equipex Union (ANR-10-EQPX-52-01), the Labex NIE projects (ANR-11-LABX-0058 NIE) and CSC (ANR-10-LABX-0026 CSC) within the Investissement d'Avenir program ANR-10-IDEX-0002-02, the International Center for Frontier Research in Chemistry (icFRC).

## Supporting Information

## Coherent Coupling of WS$_2$ Monolayers with Metallic Photonic Nanostructures at Room Temperature:

Shaojun Wang,[†,#] Songlin Li,[†,#] Thibault Chervy,[†,#] Atef Shalabney,[†,‡] Stefano Azzini,[†]

Emanuele Orgiu,[†] James A. Hutchison,[†] Cyriaque Genet,[†] Paolo Samorì,[†] and Thomas W. Ebbesen[†*]

[†] ISIS & icFRC, University of Strasbourg and CNRS, Strasbourg 67000, France

[‡] Braude College, Snunit St 51, Karmiel 2161002, Israel

*Email: ebbesen@unistra.fr

[#]These authors contribute equally for this work.

1. **Scanning electron microscope (SEM) image of the hole array sample**

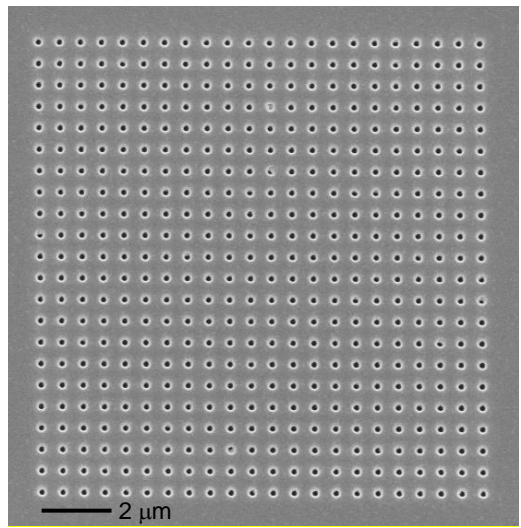

**Figure S1.** (a) SEM image of the plasmonic hole array. The array, milled through a 260 thick Au film, is a square array of period $P$ = 530 nm and hole diameter $D$ = 120 nm.

## 2. Dispersion of 2D materials

In Figure S2 b,d are shown the refractive indices of the $WS_2$ and $MoS_2$ monolayers respectively, extracted from the transmission spectra in Figure S2a and c. The imaginary part of the refractive indices emphasizes the exceptional absorption characteristics of the exciton band in the 2D materials. The full details of the fitting for the visible spectra are included in Table S1 and S2.

Table S1 Monolayer $WS_2$ dispersion parameters

| *Oscillator No.* $j$ | *Oscillator strength* $f_j$ | *Resonance energy* $E_j(eV)$ | *Damping constant* $\Gamma_j(eV)$ |
|---|---|---|---|
| 1 | 1.59 | 2.0195 | 0.028 |
| 2 | 0.70 | 2.2379 | 0.20 |
| 3 | 2.95 | 2.4087 | 0.15 |
| 4 | 2.80 | 2.5996 | 0.30 |
| 5 | 12 | 2.850 | 0.23 |

Table S2 Monolayer $MoS_2$ dispersion parameters

| *Oscillator No.* $j$ | *Oscillator strength* $f_j$ | *Resonance energy* $E_j(eV)$ | *Damping constant* $\Gamma_j(eV)$ |
|---|---|---|---|
| 1 | 0.65 | 1.9001 | 0.040 |
| 2 | 0.25 | 1.9315 | 0.050 |
| 3 | 1.2 | 2.0516 | 0.080 |
| 4 | 5 | 2.3065 | 0.8 |
| 5 | 12 | 2.4 | 1 |
| 6 | 24 | 2.87 | 0.35 |

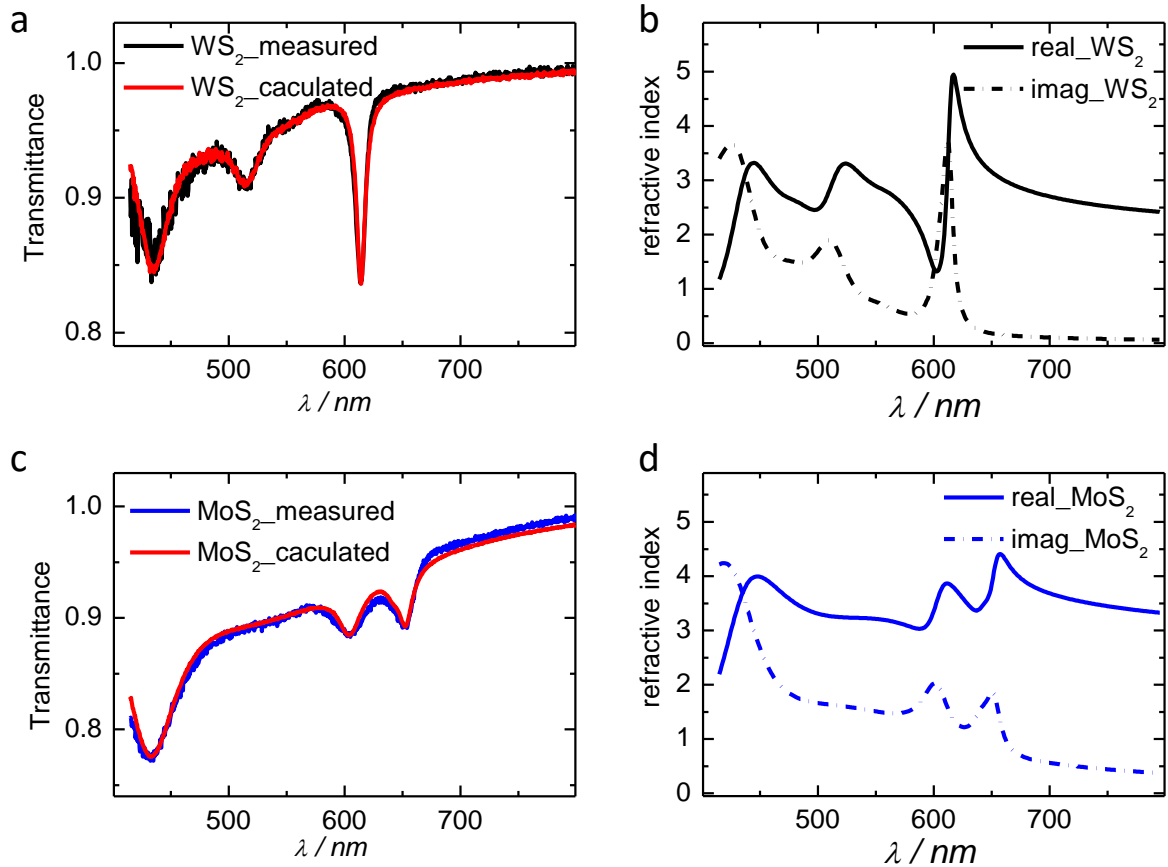

**Figure S2.** The real and imaginary refractive indices of monolayer $WS_2$ (b) and $MoS_2$ (d) calculated from the transmission spectra of the monolayer flake on quartz shown in (a) and (c) respectively.

**3. Comparing transmission, reflection of FP cavities simulated by transfer matrix method**

In order to model the spectral properties of the FP cavities, we used the standard formalism of the transfer matrix method (TMM).[1] It relies on solving Maxwell's equations at each interface of a multilayer stack, each layer being characterized by a complex refractive index. The refractive index of the silver mirrors (50 nm thick) and of the LiF spacer layers were obtained from regular data bases.[2-3] The complex refractive indices of monolayer $WS_2$ and $MoS_2$ are taken from the data in Figure S2. The thickness of the two LiF spacers were tuned to 86 nm for $WS_2$ and 92 nm for $MoS_2$ respectively and those of the $WS_2$ and $MoS_2$ monolayers were considered as 0.618 nm and 0.646 nm respectively.[4] All the parameters of the stack having been determined, the TMM was used to model the transmission/reflection spectra of $WS_2$ and $MoS_2$ cavities at normal incidence conditions. The splitting in reflection is compared to that in transmission as shown in Figure S3. The splitting in reflection $\hbar\Omega_{R\_R} = 67$ meV is slightly smaller than that in transmission $\hbar\Omega_{R\_T} = 70$ meV for the case of $WS_2$ which has very sharp *A* exciton absorption band. The predicted Rabi-splitting is smaller than that of experimental results (101 meV) likely due to errors in the assumption of the thickness of the monolayer and the refractive index of the silver mirrors. For $MoS_2$, the splitting in reflection ~ 41 meV and ~50 meV in transmission. The average splitting ratio (1.51) observed for $WS_2$ compared to $MoS_2$ cavity is comparable with the ratio of exciton transition dipole moment in each case $\sqrt{\dfrac{1.59}{0.65}} = 1.56$.

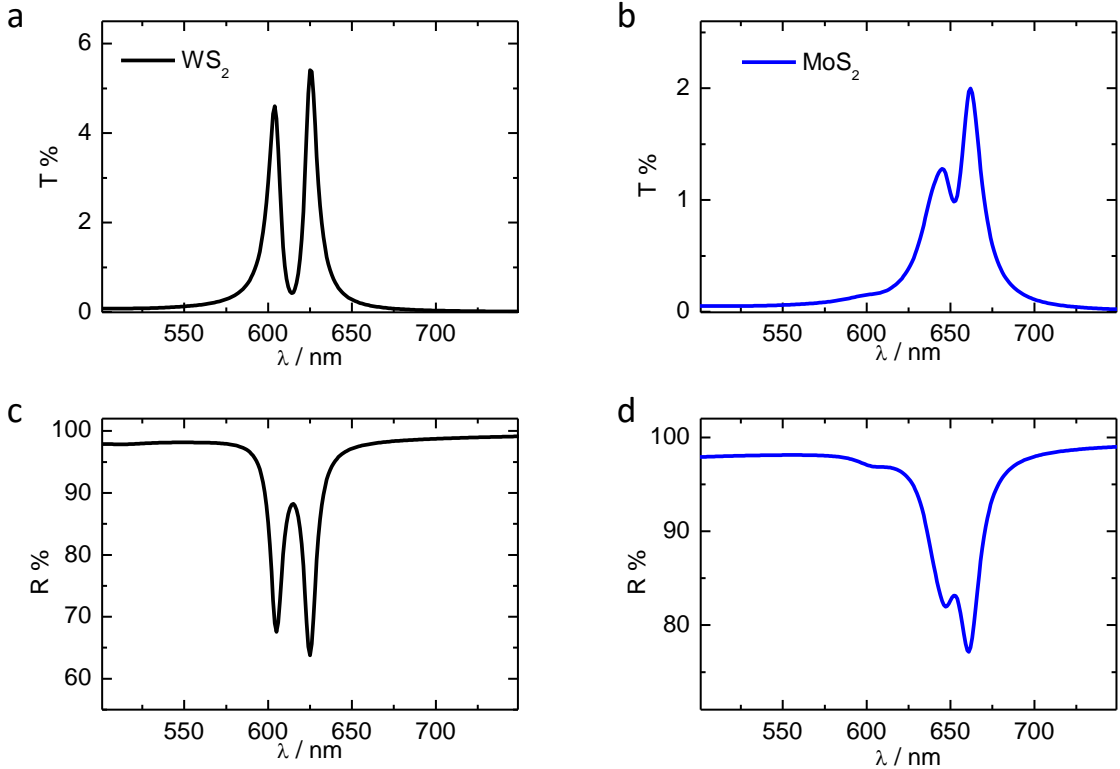

**Figure S3.** Simulated transmission (a) (b) and reflection (c) (d) of monolayer $WS_2$ (a) (c) and $MoS_2$ (b) (d) sandwiched in the middle of FP cavities tuned such that the fundamental mode is resonant with the *A* exciton transition for each case.

## 4. Mixing coefficients for strong coupling of monolayer WS$_2$ with the TE FP cavity mode

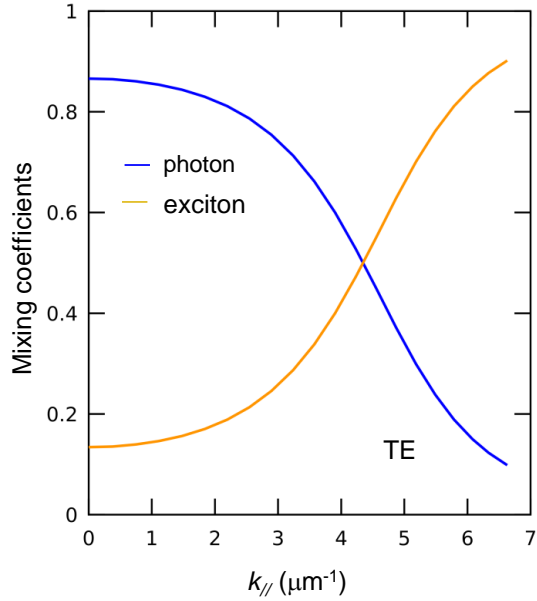

**Figure S4.** The mixing coefficients of the P- band for coupling of the WS$_2$ monolayer with the TE FP cavity mode. Blue and yellow curves represent the photonic and excitonic content of the P- band respectively.

## 5. PL spectra at resonance for the strong coupling of monolayer WS$_2$ with the TM/TE modes of the plasmonic hole array

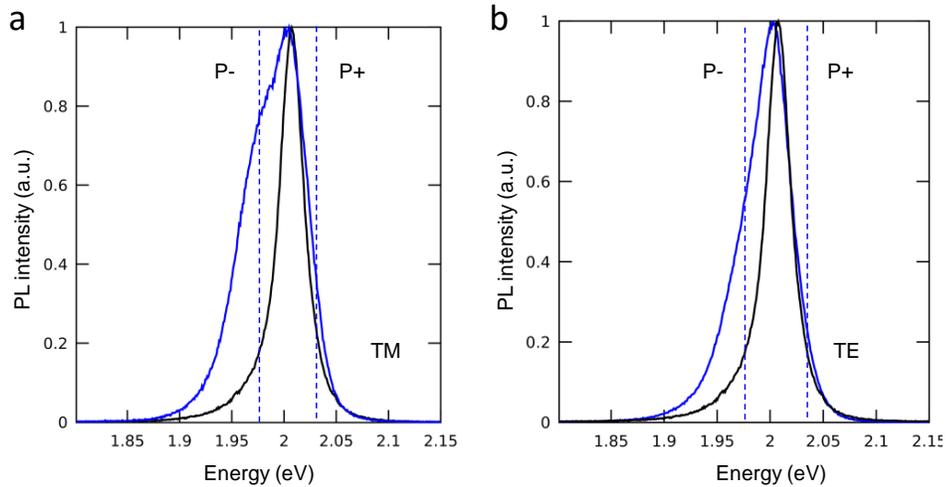

Figure S5. PL spectra (solid blue curve) of the plasmonic hole array with monolayer WS$_2$ analyzed in (a) TM and (b) TE polarization. Both spectra were obtained at the resonant condition. The black solid curve is the PL spectrum of uncoupled exciton. The vertical dashed lines represent the energy of P+/-, measured in reflection.